\documentclass[fleqn,twoside]{article}
\usepackage{espcrc2}

\usepackage{graphicx}
\usepackage[figuresright]{rotating}

\newcommand{\AmS}{{\protect\the\textfont2
  A\kern-.1667em\lower.5ex\hbox{M}\kern-.125emS}}

\newcommand{\sect}[1]{Sect.~\ref{#1}}
\newcommand{\la}{\langle}
\newcommand{\ra}{\rangle}
\newcommand{\be}{\begin{equation}}
\newcommand{\ee}{\end{equation}}
\newcommand{\bea}{\begin{eqnarray}}
\newcommand{\eea}{\end{eqnarray}}
\newcommand{\nn}{\nonumber}
\newcommand{\omk}{\omega_{\rm kin}}
\newcommand{\oms}{\omega_{\rm spin}}
\newcommand{\MeV}{\;\mbox{MeV}}
\newcommand{\GeV}{\;\mbox{GeV}}
\newcommand{\fm}{\;\mbox{fm}}
\newcommand{\itcaption}[1]{\caption{\it{#1}}}

\title{Status of heavy quark physics from the lattice}

\author{Nicolas Garron \address{DESY, Platanenallee 6, 15738 Zeuthen,  Germany}}
       
\begin{document}
\begin{abstract}
\vspace{-0.3cm}
In this short review, I present a summary of various methods used to simulate
heavy quarks on the lattice. I mainly focus on effectives theories,
and give some physical results. 
\end{abstract}

\maketitle
\vspace{-0.5cm}
\section{Introduction}
Lattice simulations provide a powerful tool to study QCD beyond perturbation 
theory. 
In particular, for quark masses around or below that of the strange quark,
today's lattices offer the possibility of very accurate and direct calculations 
of hadronic observables.
The recent feasibility of simulating dynamical fermions 
(i.e. without neglecting the effects of internal quark loops)
provide a significant 
reduction (and a better control) of the systematic errors. 
However, simulating heavy quarks on a lattice is much more challenging
than simulating the strange quark. 
This is rather unfortunate, because precise calculations
in the heavy-quarks sector are needed to constrain the standard model 
(see, e.g.~\cite{Yao:2006px,Charles:2004jd,Mackenzie:2006un,Lellouch:2002nj}). 
The problem of heavy quarks on the lattice is a long-standing one,
and various approaches have been proposed
already more than 20 years ago. Of course these ideas have grown up, 
and together with the improvement of the lattice techniques, 
they have become more and more predictive. 
In addition to that, other very interesting strategies have been proposed 
more recently. Some of them have already given physical results, and
others are not in that stage yet. 
The scope of this talk is to present the various methods  
(and their limitations) that exist to deal with heavy quarks on the lattice.
I present also a (limited) selection of results which have been obtained
by different lattice groups.
I would like to apologize to all colleagues that I cannot cite, due to 
space limitations.
For an overview of the most recent lattice results see~\cite{Onogi:latt06}.

This paper is organized as follow:
In \sect{sec:heavy}, I review the problem 
associated with simulating 
heavy quarks on a lattice. In \sect{sec:effth}, I present 
an overview of the different approaches.  
In the last section, I report some recent physical results.

\section{Heavy quarks on the lattice}
\label{sec:heavy}

Lattice QCD allows for the computation of physical quantities 
from first principle, i.e. directly from the Lagrangian of QCD.
But working in a finite and discrete space-time implies 
systematic errors, which have to be under control. 
The main problem with heavy quarks is that the discretization
errors 
(when a traditional Wilson-like fermionic action is used) 
are proportional to powers of the bare quark mass.
Thus one has to require $am_{\rm quark}\ll 1 $.
To simulate a b-quark at its physical mass ($\approx 5 {\rm GeV}$)
on a space-time of volume $(aN)^4=(2 \;{\rm fm})^4$,
one needs $N\gg 50$ points for each space-time dimension, 
which is impossible with present computers~\footnote{Direct 
simulations of the c-quark are doable, but some care has to be 
taken to control the discretization errors~\cite{Rolf:2002gu}.}.

One can simulate various quark
masses in the regime where the discretization errors are under 
control (say around the charm quark), and extrapolate the 
results to the b-quark. The problem is that this extrapolation
is done over a large range (because $m_{\rm b}\approx 4 m_{c}$),
and the associated systematic error
is then difficult to control~\footnote{This extrapolation
can be replaced by a -more safe- interpolation, by the use
of an effective theory. See~\cite{Rolf:2003mn} for a 
quenched computation of $f_{B_s}$ done in this way.}.\\
Another possibility is to use effective theories, like heavy quark 
effective theory (HQET) or non relativistic QCD (NRQCD).
Since the heavy quark mass 
is much larger than the other scales (like its 3-momentum or $\Lambda_{\rm QCD}$),
the idea is to expand the QCD Lagrangian in inverse powers of the heavy quark mass
and keep only the leading terms (I give more details below). 
The procedure results an effective theory, which has to be matched with QCD.
This matching is a source of uncertainty when it is done perturbatively.
Recently, some efforts have been made to remove the largest
discretization errors by adding appropriate terms to the 
Lagrangian. One obtains a Lagrangian which describes
both light and heavy quarks. One of the main problem with 
these methods is to compute the coefficients which come in front 
of the terms that one adds to the Lagrangian. 

In the next section, I present shortly these effective theories, and their
lattice discretization (see~\cite{Khan:2005cf} for a pedagogical review).
\section{Effective theories}
\label{sec:effth}

A number of simplifications can be made when one deals with one or several 
heavy quarks. 
To make this explicit, one can derive an effective Lagrangian, which 
is much simpler than the original.
The starting point is the observation that the momentum of a heavy quark 
inside a hadron can be written as $p=m_Qv+k$. In that decomposition, 
$v$ is the velocity, and $k$ the residual momentum, 
which is zero if the heavy quark is on-shell. 
For a heavy quark interacting with light degrees of freedom, the components 
of the residual momentum $k$ are of order $\Lambda_{\rm QCD}$, and therefore
much smaller than  $m_Q$.
One can separate the higher and lower components (this refers to the case
where the $\gamma$ matrices are written in the Dirac basis)
of the heavy quark field
$Q(x)$ in 
$ h_\pm(x) = e^{im_Qv.x}P_\pm Q(x)$, where
$P_\pm={1\pm v\hspace{-0.15cm}\slash \over 2}$.
Then one finds the effective tree-level 
Lagrangian~\footnote{In principle one can also add
a mass term, but it can be absorbed in the definition
of the quark mass.}
(see e.g.~\cite{Neubert:2000hd} for a simple derivation) which reads 
\begin{eqnarray}
\label{eq:leff}
\lefteqn{{\cal L}_{\rm eff} = } \\
& &
 \bar h_+(x) \left[ iv.D + {(iD_\perp)^2 \over 2 m_Q} 
+{g\sigma.G \over 4m_Q} + ...\right] h_+(x) \nn
\end{eqnarray}
where the ellipse represent higher order terms.

The values of the coefficients given in eq. (\ref{eq:leff}) 
only hold at tree level~\footnote{In regularization scheme which preserves
reparametrization invariance, it can be shown that 
$\omega_{\rm kin}=1$ 
for a certain choice of the definition of 
the mass}.
They have to be renormalized to include loops effects.
This is usually done though a perturbative matching
with QCD. One computes a physical process at a certain order of
the effective theory, and imposes the result to be equal to
its QCD value. 
However, on the lattice, 
a perturbative computation of these coefficients
will lead to a results which is power divergent in the continuum limit.
Thus a non-perturbative matching with QCD is needed,
this has been done in the case of
HQET~\cite{Heitger:2003nj}(I will give an explicit
example in~\sect{sec:mb}), but not for NRQCD.

\subsection{Heavy quark effective theory}
HQET describes a situation where only one heavy quark is present, 
like heavy-light mesons. 
It is then more natural to write everything in the rest frame 
of this heavy quark. 
The previous effective Lagrangian is seen as
an expansion in $\Lambda_{\rm QCD}/m_Q$. 
With $\Lambda_{\rm QCD}\sim 500$ MeV and $m_Q\sim 5$ GeV 
the theory is expected to have roughly a $10 \%$ precision
at the leading order, and $1\%$ at the next to leading order.
At the leading order, 
when $m_Q\to \infty$, 
the Lagrangian is then
${\cal L}_0= \bar h_+(x) \left[ iD_0 \right] h_+(x)$ .
It represents a heavy quark acting only as a static color source.
One can see that, at this order, the light quarks are independent 
of the flavor and of the spin of the heavy quark. 
The second and the third terms appear at the next to 
leading order, and represent respectively the interaction due to the 
motion and to the spin of the heavy quark :
\begin{eqnarray}
\label{eq:lkin}
{\cal L}_{\rm kin}&=&-\bar h_+(x) \left[{{\vec D}^2 \over 2 m_Q}\right] h_+(x) 
\\
\label{eq:lspin}
{\cal L}_{\rm spin}&=&-
\bar h_+(x)\left[ {g{\vec \sigma}.{\vec B} \over 4m_Q} \right] h_+(x) .
\end{eqnarray}
A lattice formulation of HQET at the leading order is the so-called
Eichten-Hill action~\cite{Eichten:1989zv}. 
It is important to note that the higher order terms
appear only as insertion in the static green functions. 
Explicity, at the $1/m_Q$ order, 
one writes
\begin{eqnarray}
\lefteqn{
\exp\left( - {\cal S}_{\rm light} 
           - {\cal S}_{\rm HQET} \right)
= \exp \left( - {\cal S}_{\rm light} 
              - {\cal S}_{\rm stat} \right)
}
\nn\\
&&\times
 \left(1 + a^4\sum_{x} 
\left[\omk {\cal O}_{\rm kin}+\oms {\cal O}_{\rm spin} \right]\right)
\;.
\nn
\nopagebreak
\end{eqnarray}
For a green function of an operator $\cal O$, this means:
\begin{eqnarray}
\lefteqn{\la {\cal O} \ra = \la {\cal O} \ra_{\rm stat} 
+\omk a^4\sum_x\la {\cal O}{\cal O}_{\rm kin} (x) \ra_{\rm stat}
}\nn\\
&&
+\oms a^4\sum_x\la {\cal O}{\cal O}_{\rm spin}(x) \ra_{\rm stat}
\nn \;, 
\end{eqnarray}
where $\la {\cal O} \ra_{\rm stat}$ is the expectation value 
of $\cal O$ given by the Lagrangian 
${\cal L}_{\rm light} +{\cal L}_{\rm stat}$. 
In that way, the continuum limit is well defined, and the theory 
is (power-counting) renormalizable. 
In the previous expressions $\cal O_{\rm kin}$ and 
${\cal O}_{\rm spin}$ are the lattice version of the operators 
corresponding to (\ref{eq:lkin}) and (\ref{eq:lspin}) respectively.
The coefficients $\omk$ and $\oms$ are fixed by the matching with QCD.

\subsection{Non relativistic QCD}
In a heavy-heavy hadron, the dynamics is rather different than 
with heavy-light, and HQET is not an appropriate theory.
Starting from the same effective Lagrangian~(\ref{eq:leff}),
it is possible to derive another effective theory, called 
NRQCD\footnote{NRQCD can also be used for heavy light
mesons}~\cite{Thacker:1990bm}.
In a quarkonium, the typical momentum of a heavy quark should be 
such that its kinetic and potential energy are balanced  
$\la {{\vec p}^2\over m_Q}\ra \sim \la {\alpha_s \over r} \ra $. 
Since the distance $r$ between the quark and the antiquark is of 
order $1/|{\vec p}|$, one find that $\la |\vec p| \ra \sim \alpha_s m_Q$.
Then a natural separation of the scales is given by $|\vec v|\sim\alpha_s$ : 
$$ m_Q \gg |\vec p|\sim m_Q|\vec v| \gg {{\vec p}^2\over m_Q} \sim m_Q {\vec v}^2 \gg ...$$
Therefore in NRQCD, the power counting is different from that of 
HQET~\cite{Thacker:1990bm,Lepage:1992tx},
the two first terms of the effective Lagrangian are of order
$m_Q v^2$ and the third one (the 'spin' term) is of order $m_Q v^4$.
The physical picture implies that the kinetic term
cannot be treated as a small $1/m_Q$ correction (as in HQET), but has 
to be included already at the leading order.
On the lattice, this implies that the continuum limit cannot be taken. 

The authors of~\cite{Lepage:1992tx} gave an improvement program
to reduce the errors below $10\%$. 
The principle is to add
next to leading
order corrections to this Lagrangian. 
First there are some relativistic corrections ($\vec v^2\sim 0.1$
for the b-quark, and $0.3$ for the c-quark). 
Then the leading discretization errors are of order $O(a^2 {\vec p}^2)$
and $O(a{{\vec p}^2\over 2m_Q})$, which are the same
order of magnitude as the relativistic corrections. 
These corrections to the Lagrangian introduce new coefficients, 
which contain power law divergences (in $g^2/am_Q$), because of the 
non-renormalizability of the theory. 
They are computed in perturbation theory, implying that
the lattice spacing should not be too small (typically $a\sim 1/m_Q$).
To reduce the lattice spacing (and increase the precision),
one should add more and more terms to the Lagrangian.
Naively, these radiative corrections can contribute for $20$ or $40\%$ errors,
so these perturbative calculations with care. 
A lot of work has been done is the last years to obtain
lattice results for the hadron spectrum with a very
high accuracy, and the inclusion of dynamical fermion
plays a crucial role, as discussed in \sect{sec:unq}.

\subsection{Relativistic heavy quarks}
\label{sec:relhq}

The lattice formulations of both HQET and NRQCD are obtained 
from their continuum version. One first writes down
an effective Lagrangian in an euclidian space-time at a certain order,
then this Lagrangian is discretized. Of course the resulting
theory is only valid for heavy quarks (and low velocity for
NRQCD).

This is rather different in the Fermilab approach 
(see e.g.~\cite{El-Khadra:1996mp,Kronfeld:2002pi,Kronfeld:2003sd}).
The aim is to find a lattice Lagrangian able to describe both small 
and large masses. 
The idea is to use on-shell Symanzik improvement, 
treating both $a$ and $1/m_Q$ as short distances, 
in such a way that the theory still makes sense when $am_Q>1$. 
Starting from the standard Wilson-Dirac operator, 
one can write 
\begin{equation}
\label{eq:lagfermilab}
{\cal L}^{\rm quarks}_{\rm lat}={\cal L}_{\rm Wilson}
+\sum_{\cal O}a^{{\rm dim} {\cal O}-4} c_{\cal O} {\cal O}_{\rm lat} .
\end{equation}
The (dimension 5) kinetic and spin operators appear in ${\cal O}_{\rm lat}$,
in the the previous expression. 
Their coefficients $c_{\cal O}$ are functions of $am_Q$ which are tuned 
to reproduce their continuum values (taken e.g. from NRQCD). 
Unfortunately this matching 
is done done (at least partially) in perturbation theory.
The claim is that when $a\to 0$, the action reduces to the 
(improved) Wilson action. Then contrary to NRQCD, the continuum limit 
of~(\ref{eq:lagfermilab}) exists. 
Once all the coefficients have been computed
at a sufficient order, the Lagrangian ($\ref{eq:lagfermilab}$),
has the properties of NRQCD (or HQET) for $am_Q>1$, and reduces to the light
action for $am_Q<1$. 
In the same spirit, the authors of ~\cite{Aoki:2001ra} 
reduce the discretizations errors up to order $a\Lambda_{\rm QCD}$, 
by using 4 dimension 5 operators (with 4 different coefficients). 
In a very recent work~\cite{Christ:2006us}, it is shown 
that the coefficients of these operators are not independent,
and that, indeed, all errors of order $O(a|\vec p|)$ and 
$O((am_Q)^n)$ can be removed (for arbitrary n) by a proper choice
of 3 coefficients (including the bare quark mass $m_0$). 
Moreover, contrary to NRQCD or to the Fermilab method, 
these coefficients can be computed 
non-perturbatively~\cite{Lin:2006ur}, and then would represent a 
great improvement.

\section{Physical results}
\subsection{Unquenched results}
\label{sec:unq}

\begin{figure}[!htb]
\begin{center}
\includegraphics[width=7.5cm]{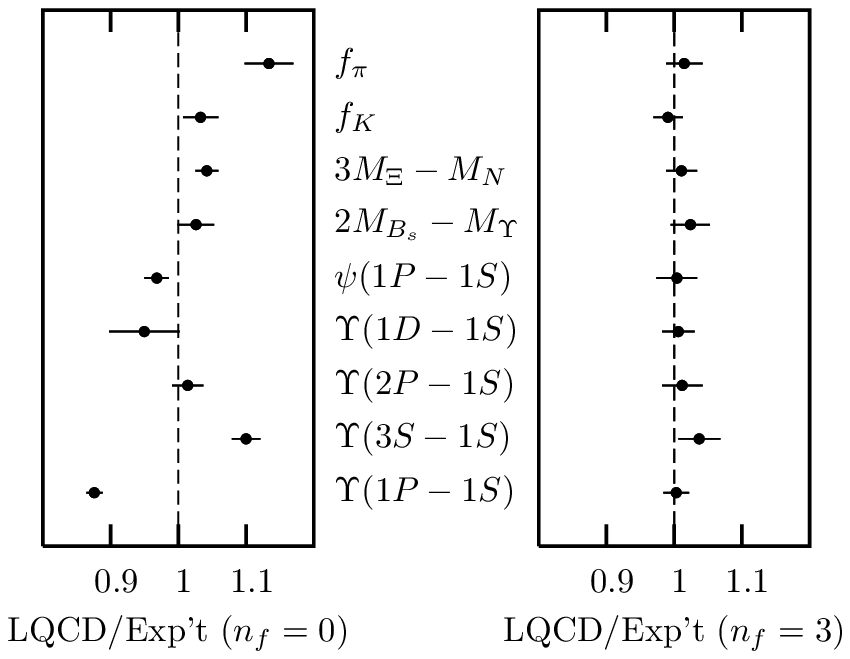}
\vspace{-1.2cm}
\itcaption{Comparison of quenched and unquenched results
with experiments.}
\vspace{-0.8cm}
\label{fig:comparison}
\end{center}
\end{figure}

\begin{figure}[!htb]
\begin{center}
\includegraphics[width=6cm]{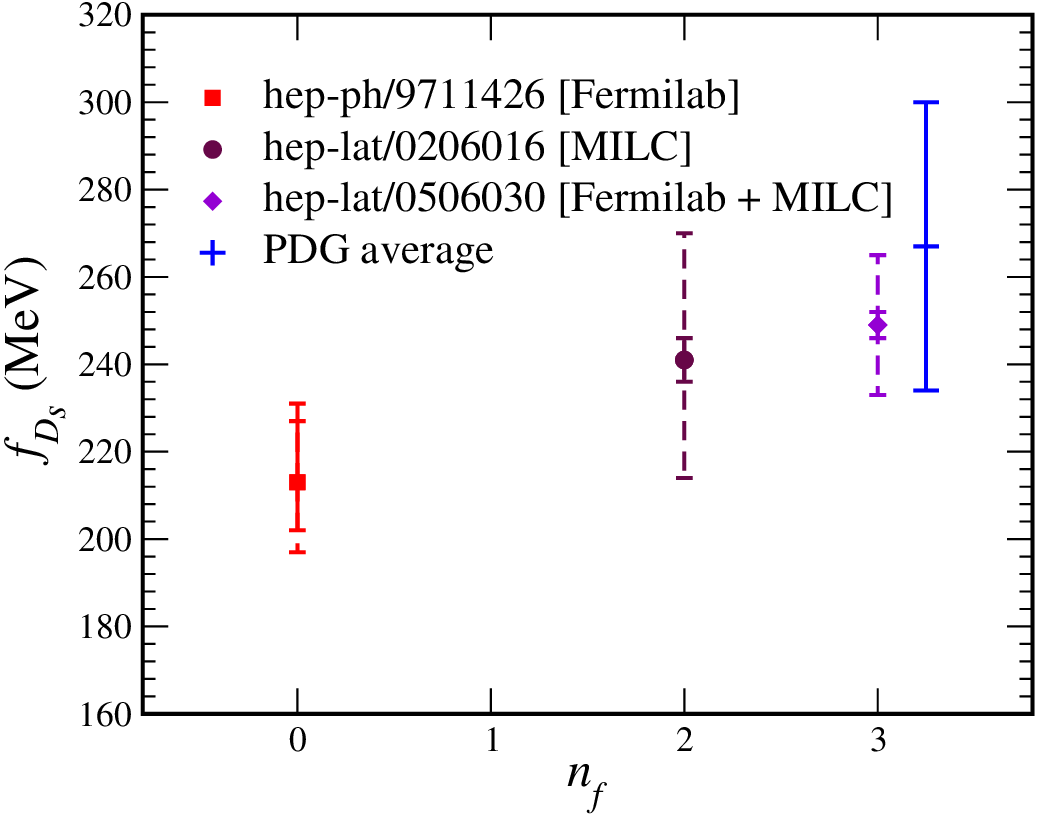}
\vspace{-1cm}
\itcaption{Comparison of quenched ($n_f=0$) and
unquenched ($n_f=2,3$) results of $f_{D_s}$ 
with its experimental value.}
\vspace{-1.cm}
\label{ffig:fDs}
\end{center}
\end{figure}

A few years ago, in a common paper~\cite{Davies:2003ik}, 
the MILC, HPQCD and Fermilab collaborations presented a 
comparison of lattice computation of hadronic observables with 
the experimental results. In the quenched approximation, the results 
agree within $10 \%$, and when the effects of quark
vacuum polarization are included, the agreement increases
to $3 \%$, as one can see in Fig~\ref{fig:comparison}.
\vspace{-0.6cm}
Last year, the same collaborations achieved the computation 
of various quantities which were experimentally unknown or poorly known 
at that time, like the decay constant of the $D_s$ and the $D_+$ mesons,
or the mass of the $B_c^+$ mesons. After precise experimental measurement
of these quantities, the main conclusion is again that the inclusion of dynamical 
fermions is very important for the agreement between lattice and
experimental results~\cite{Kronfeld:2005fy} (see e.g. Fig~\ref{ffig:fDs}).
The b-quark was simulated with a $O(a^2, v^4)$-improved NRQCD action,
and the c-quark with the Fermilab action. For the light dynamical fermions,
the staggered action was used. The problem with these fermions, is the use
of fourth root trick, to eliminate the non-physical 'tastes'. 
Since this can introduce non-locality, this
can be potentially dangerous (for a recent review about this, 
see~\cite{Sharpe:2006re}). The advantage of using staggered fermion
resides in the fact that they are numerically very cheap. Simulating
light quarks is then easier than than with other actions 
(in this work the sea quark masses go down to $\sim m_s/6$).

\subsection{Quenched results in HQET}
I report here some calculations in the b-sector, 
which present some theoretical 
advantages, despite the use of the quenched approximation.
\\

A computation of the bag parameter of ${B_s}-{\bar {B_s}}$
mixing in the static limit has been presented at this 
conference~\cite{Blossier:thisproc}.
The main advantage of this computation is that the light
quark is simulated with the overlap action~\cite{Neuberger:1997fp}, 
which presents a chiral symmetry at finite lattice spacing.
This simplifies the renormalization of the 4-quark operators,
and reduces the systematic errors. The result is 
$B_{B_s}^{\rm {\overline MS}}=0.92(3)$\footnote{Here and in the
following, the errors do not take into account the use of the 
quenched approximation.}.\\

The two last exemples concern the mass of the b-quark and 
the decay constant $f_{B_s}$.
One common point is the use of a small volume 
(here of space extent $L_1\sim 0.4$ fm) 
to simulate a relativistic b-quark with discretization 
errors under control.
The authors of~\cite{Guazzini:2006bn} have  
used HQET together with the step scaling 
method, to compute these 
quantities~\cite{deDivitiis:2003iy,deDivitiis:2003wy}. 
One starts by the computation of
a finite-volume observable $\Phi(L_1)$. 
The evolution to a larger volume
$L_{\rm i+1}=kL_{\rm i}$ is given by a step scaling function (ssf), 
defined by $\sigma_(L_2)={\Phi(L_2) / \Phi(L_1)}$.
If the volume of space extent $L_\infty=L_{\rm final}$ is large enough 
to be considered as an infinite volume, one has
\begin{equation}
\label{eq:evol}
\Phi(L_\infty) = \sigma(L_{\rm final}) ... \sigma(L_3)\sigma(L_2)\Phi(L_1)
\end{equation}
In the r.h.s. of (\ref{eq:evol}), $\Phi(L_1)$ is computed in QCD, 
and the ssf are interpolated 
to the b-quark mass by using QCD with lower masses together 
with a static calculation.
The l.h.s can be an HQET prediction
or an experimental input. 
One choice of observable is the pseudo-scalar mesons mass.
In the infinite volume, it is fixed to the experimental value 
of the $B_{s}$ mass. The  b-quark mass is then extracted
from the r.h.s of~(\ref{eq:evol}).
Of course, in the case of the decay constant, the l.h.s. of 
(\ref{eq:evol}) is a prediction. 
The 'static' results are 
$m_{\rm b}^{\rm {\overline MS}}(m_{\rm b})=4.421(67)\GeV$ and
$F_{B_s}=191(6)\MeV$.\\
\label{sec:mb}

A non perturbative calculation of the b-quark mass including
the $1/m_Q$ terms has been done in~\cite{DellaMorte:2006cb}.
This is a direct application of the method introduced 
in~\cite{Heitger:2003nj}, where a calculation of the b-quark
mass in the static approximation is also performed.
Here I just give a quick overview of the strategy.
For simplicity, I start with the static case.
At this order the 
meson mass is given in terms of the bare b-quark mass by
\be
\label{eq:mbav}
m_{\rm B} = m_{\rm bare}+E^{\rm stat}
\ee
We can define an observable $\Phi(L)$ both in QCD and
in HQET, such that 
$\Phi(L_\infty)=m_{\rm B}$. 
In a finite volume $L_1$, the
equivalent of (\ref{eq:mbav}) is
\be
\label{eq:match}
\Phi(L_1,M_{\rm b})^{\rm QCD}=m_{\rm bare}+\Gamma^{\rm stat}(L_1) \; 
\nn
\ee
where $\Gamma^{\rm stat}(L_\infty)=E^{\rm stat}$.
Since the bare parameters are independent of the volume,
one can use~(\ref{eq:match}) to substitute 
$m_{\rm bare}$ in~(\ref{eq:mbav}).
This gives an expression where 
$\left[E^{\rm stat}-\Gamma^{\rm stat}(L_1)\right]$ appears.
The terms $E^{\rm stat}$  and  $\Gamma^{\rm stat}$ contain 
divergences linear in the inverse lattice spacing, but they 
cancel in the difference. However, since $L_\infty$ is much larger
than $L_1$,
it is in practice very difficult 
to find a lattice spacing
common to these volumes. Instead, one can introduce an intermediate
volume $L_2=2L_1$, and the step scaling function 
\be
\sigma^{\rm stat}=L_2\left[\Gamma^{\rm stat}(L_2)-\Gamma^{\rm stat}(L_1)\right]\; .
\nn\ee
Then one can write:
\bea
\nn
m_{\rm B}=\Phi(L_1,M_{\rm b})^{\rm QCD} + 
\left[E^{\rm stat}-\Gamma^{\rm stat}(L_2)\right]+ {\sigma^{\rm stat}\over L_2}
\nonumber
\nn
\eea
Now the cancellations of the divergences occur in 
$E^{\rm stat}-\Gamma^{\rm stat}(L_2)$ and in $\sigma^{\rm stat}$.
Of course the procedure can be repeated, but it appears
sufficient in practice to do the computation with 3 different 
volumes $L_\infty\sim 1.4 \fm > L_2 \sim 0.7 \fm > L_1 \sim 0.35 \fm$.
Finally, the mass of the B meson is fixed to its experimental value
and one extracts the RGI b-quark mass by an interpolation
($\Phi(L_1,M_{\rm b})$ has to be computed for a few quark masses
around the b-quark mass). 
The whole procedure can be generalized at the next order of HQET. 
The starting point is to rewrite~(\ref{eq:mbav}) incuding 
the $1/m_Q$ term~\footnote{In principle there should be a term 
$\omega_{\rm spin} E^{\rm spin}$, but it can be eliminated by 
considering a spin-average B meson 
$ m_{\rm B}^{\rm av}={1\over 4}( m_{\rm B}+3m_{{\rm B}^*})$.
Then the spin-splitting becomes 
a separate issue.}
\bea
m_{\rm B}^{\rm av}=
m_{\rm bare}+E^{\rm stat}+\omega_{\rm kin}E^{\rm kin}
\nn\eea
Then one has to find an observable to eliminate 
$\omega_{\rm kin}$, and applies the same techniques 
than in the static case (e.g. two other ssf are needed).
The result is $m_{\rm b}^{\rm {\overline MS}}(m_{\rm b}) =4.437(48)\GeV$.\\
This non-perturbative method presents several good features.
It is based on general hypotheses, and so can be applied to various quantities, 
like heavy-light decay constants. The cost is rather moderate, so 
it should be possible to generalize it to full QCD.
Since the problem of power-law divergences is cured, the theory is
well-defined in the continuum limit. 
Because calculations can be done at the NLO of HQET
(the NNLO corrections are expected to be very small),
a very good precision can be reached. 
\section{Outlook}

Heavy flavor physics on the lattice is a very active field,
and lot of progress has been achieved recently. Some of 
them are technical (all-to-all propagators~\cite{Foley:2005ac}, 
noise reductions~\cite{DellaMorte:2005yc}, 
new algorithms), and other are conceptual (new strategies,
new effectives actions, non-perturbative renormalization).
Results with simulations done with dynamical quarks
are already available. Although a lot of them are done with
the controversial staggered quarks, simulation of dynamical 
fermions with other actions are on their way. 
They have a major role to play in constraining the 
standard model and in the search for new physics.

{\bf Acknowledgments.}\\
I thank D. Guazzini, L. Lellouch, O. P\`ene, 
A. Shindler, R. Sommer for useful discussions.
I would also like to thank L. Lellouch and R. Sommer
for a careful reading of the manuscript. 
\bibliographystyle{h-elsevier}
\bibliography{mybib}
\end{document}